\documentclass[final,3p,authoryear]{elsarticle}

\usepackage{amsmath,amsfonts,mathrsfs}
\usepackage{epsfig,graphicx}
\usepackage{multirow,longtable}
\usepackage{array}
\usepackage{placeins}
\usepackage{color}
\usepackage{lineno,hyperref}
\usepackage{natbib}
\usepackage{xcolor}

\graphicspath{{./Figures/}}

\begin{document}

\begin{frontmatter}

\title{Attention decay in science}

\author[aalto]{Pietro Della Briotta Parolo}
\author[aalto]{Raj Kumar Pan}
\author[bosch]{Rumi Ghosh}
\author[hp]{Bernardo A. Huberman}
\author[aalto]{Kimmo Kaski}
\author[aalto]{Santo Fortunato}
\address[aalto]{Complex Systems Unit, Aalto University School of Science, P.O.  Box 12200, FI-00076, Finland} 
\address[bosch]{Robert Bosch LLC, Palo Alto, CA 94304, USA} 
\address[hp]{Mechanisms and Design Lab, Hewlett Packard Enterprise Labs, Palo Alto, California, USA} 

\begin{abstract}
The exponential growth in the number of scientific papers makes it increasingly difficult for researchers to keep track of all the publications relevant to their work. Consequently, the attention that can be devoted to individual papers, measured by their citation counts, is bound to decay rapidly. In this work we make a thorough study of the life-cycle of papers in different disciplines. Typically, the citation rate of a paper increases up to a few years after its publication, reaches a peak and then decreases rapidly. This decay can be described by an exponential or a power law behavior, as in ultradiffusive processes, with exponential fitting better than power law for the majority of cases. The decay is also becoming faster over the years, signaling that nowadays papers are forgotten more quickly. However, when time is counted in terms of the number of published papers, the rate of decay of citations is fairly independent of the period considered. This indicates that the attention of scholars depends on the number of published items, and not on real time.
\end{abstract}

\end{frontmatter}


\section{Introduction}
Scientific publications in peer reviewed journals serve as the standard medium through which most of the progress of science is recorded. Besides offering a mechanism for claiming priorities and exposing results to be checked by others, publishing is also a way to attract attention of other scientists working on related problems. Attention, measured by the number and lifetime of citations, is the main currency of the scientific community, and along with other forms of recognition forms the basis for promotions and the reputation of scientists~\citep{petersen_reputation_2014}. As Franck~\citep{franck_scientific_1999}, Klamer and van Dalen~\citep{klamer_attention_2002} have pointed out, there is an attention economy at work in science, in which those seeking attention through the production of new knowledge are rewarded by being cited by their peers, whose own standing is measured by the amount of citations they receive.

The attention economy is also at work in many other fields besides science, ranging from entertainment to marketing, and is responsible for the phenomenon of stars, i.e., people whose income in attention far exceeds the norm in their own endeavors. Moreover, attention is a strong motivator of productivity. Recently, it has been shown that the productivity of YouTube videos exhibits a strong positive dependence on the attention they receive, measured by the number of downloads~\citep{huberman_crowdsourcing_2009}. Conversely, a lack of attention leads to a decrease in the number of videos uploaded and the consequent drop in productivity, which in many cases asymptotes to no uploads whatsoever.

Decision making and marketing, among others, are based on the mechanisms ruling how attention is stimulated and maintained~\citep{kahneman73,pashler98,pieters99,dukas04,reis06}. Over the past years, thanks to the Internet, a huge amount of data has allowed a thorough investigation of the dynamics of collective attention to online content, ranging from news stories~\citep{dezso06,wu_novelty_2007,ghosh14}, to videos~\citep{crane08} and memes~\citep{leskovec09,matsubara12,weng12}. Here attention is measured by the number of users' views, visits, posts, downloads, tweets. It is also noted that the attention decays over time, not only because novelty fades, but also because the human capacity to pay attention to new content is limited. A typical temporal pattern is characterized by an initial rapid growth, followed by a decay. The decay turns out to be slower than exponential: power law fits give the best results, stretched exponentials being preferable in particular cases~\citep{wu_novelty_2007}.

In this paper we focus on the decay of attention in science, on the basis of scientific articles, which like any other content, become obsolete after a while. Typically this happens because their results are surpassed by those of successive papers, which then ``steal'' attention from them. The problem of the obsolescence of scientific contents has received a lot of attention in scientometrics. The typical approach is to study the evolution of the number of citations received by a paper in a given time frame (usually one year), since its publication. The nature of the decay has been controversial, between claims of an exponential trend~\citep{avramescu79,nakamoto88, medo11} and analyses supporting a slower power law curve~\citep{pollmann00,redner05,bouabid11,bouabid13}.  This is partly due to the different types of analysis and the use of distinct data sources. Note that patterns of individual papers are usually noisy, as one cannot count on the high statistics available for online contents: the number of tweets posted on a single popular topic may exceed the total number of scientific publications ever made.

On the other hand, in contrast to online sources, bibliographic databases enable one to perform a longitudinal study of the life cycles of papers. In this work we make a systematic analysis of papers' life cycles, across different scientific fields and historical periods. We find that the decay of attention for individual papers can be described both by exponential and power law behaviors. Exponential fits turn out to be preferable in the majority of cases. These results are compatible with a relaxation of attention modeled by ultradiffusion, as observed for the popularity of online content~\citep{ghosh14}.  We also found that attention is dying out more rapidly with time. However, due to the ongoing exponential growth of scientific publications, which is known to influence citation patterns \citep{Egghe2000,Yang2010}  , we conjecture that the faster decay observed nowadays is a consequence of the much larger pool of papers among which attention has to be distributed. In fact, if time is renormalized in terms of the number of papers published in the corresponding period (e.g., in each given year), we find that the rescaled curves die out at comparable rates across the decades.

\section{Material and methods}
\subsection{Data description}
Our data set consists of all publications (articles and reviews) written in English till the end of 2010 included in the database of the Thomson Reuters (TR) Web of Science. For each publication we extracted its year of publication, the subject category of the journal in which it is published and the corresponding citations to that publication. Based on the subject category of the journal (determined by TR) of the publication, the papers were categorized in broader disciplines such as Physics, Medicine, Chemistry and Biology (see Table 1). Most analyses are carried out using the top 10\% papers (based on their total number of citations), as it allows to include a sufficient number of papers from older times, but still keeping the number of yearly citations large enough to allow for a statistically valid analysis. The analysis of papers with relatively lower citations follow qualitatively similar behavior and is shown in the Appendix. 

\begin{table}[b]
\centering
\label{tab:stats}
\caption{Basic statistics of the different scientific fields we considered: Clinical Medicine, Molecular Biology, Chemistry and Physics. They represent the most active fields in terms of the total volume of publications. Here, $N_\mathrm{P}$ is the number of publications in a given field, $c_{\mathrm{max}}$  is the maximum number of citations to a given paper in that field and $\langle c \rangle$ is the average number of citations to all the papers in that field.}  
\vspace{0.3cm}
\begin{tabular}{llll}
\hline
Field & $N_\mathrm{P}$ & $c_{\mathrm{max}}$ & $\langle c \rangle$ \\ 
\hline
Clinical Medicine & 10833626 & 25604 & 11  \\ 
Molecular Biology & 2849144 & 296498 & 24  \\ 
Chemistry & 4565197 & 134441 & 14  \\ 
Physics & 5583183 & 31759 & 13  \\ 
\hline
\end{tabular}
\end{table}

\subsection{Data fitting and F-statistics}
We measure the trend in the temporal evolution of the different plots using the least square method. We consider the F-statistics for a significant linear regression relationship between the response variable and the predictor variable. We used it to compare the statistical models that best fit the population from which the data were sampled. As the F-score takes into account both the number of data points available for the fit and the number of degrees of freedom of the model, it is possible to compare the accuracy of the fit for different models with different parameters or between data sets of different size.

\begin{figure}[tb!]
\centering 
\includegraphics[width=0.5\columnwidth]{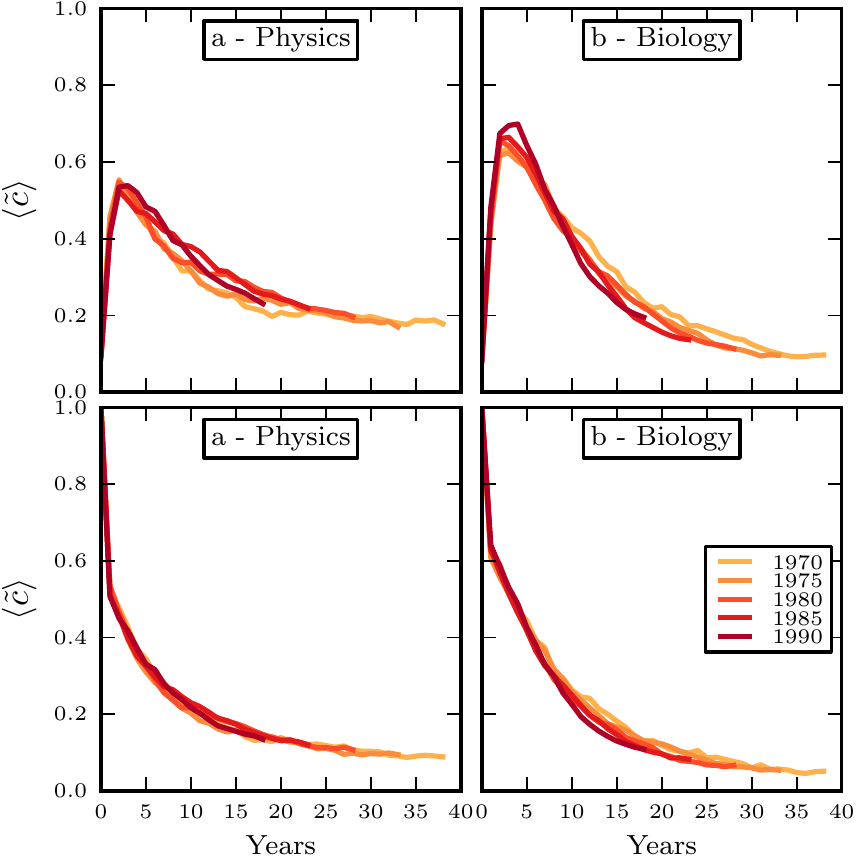}
\caption{The citation life-cycle is both field dependent and time dependent. (Top) Normalized number of citations per year received by papers in Physics and Biology published in the same year, for different publication years. Normalization is done by dividing the number of citations by the peak value reached by the paper. (Bottom) The decay in the (normalized) citation trajectory of papers in both fields after the peak year. For both disciplines, the averaged citation trajectories are calculated for papers in the top decile (top 10\%) based on their total number of citations.}
\label{fig:citation_trajectory}
\end{figure}

\begin{figure}[tb!]
\centering 
\includegraphics[scale = 1]{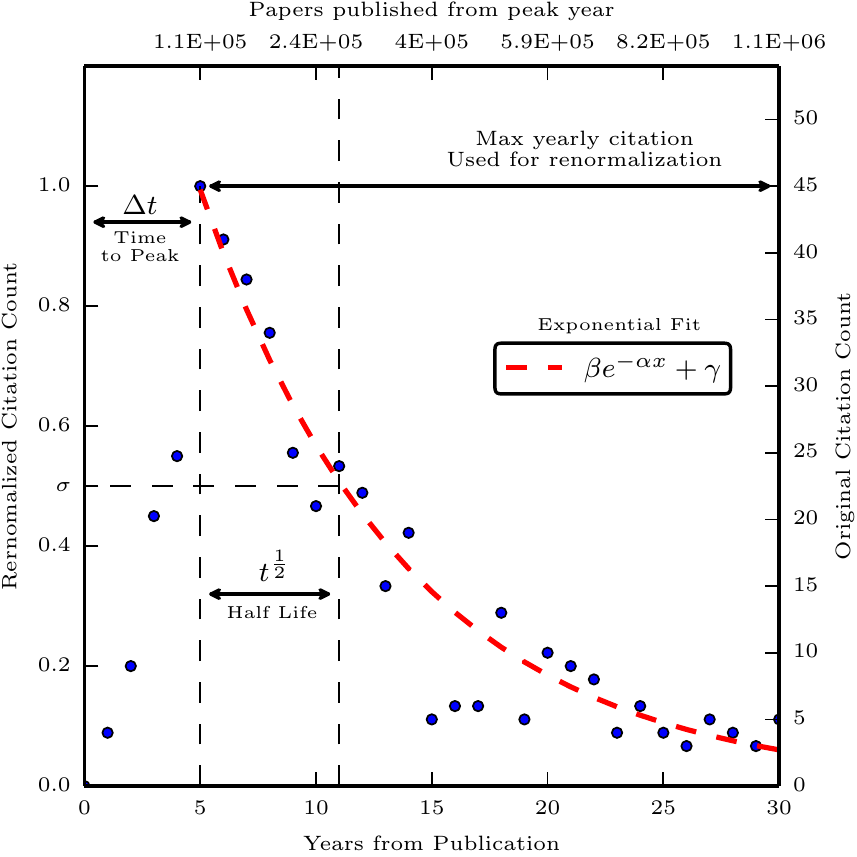}
\caption{Schematic representation of the citation evolution of a typical
paper.} 
\label{fig:individual}
\end{figure}

\section{Results and discussions}
\subsection{Evolution of the number of citations}
We first look at the way citations received by a paper change with time. Since different scientific fields are 
characterized by different volumes of publications and citations, many features of the citation trajectory are 
field dependent. However, for most fields the number of yearly citations $c_i(t)$ to a given paper $i$ rises 
after its publication and peaks within 2-7 years. The peak is followed by a decay in the number of citations 
that reflects the obsolescence of older knowledge. Fig.~\ref{fig:citation_trajectory} (top panels) shows the normalized 
citation trajectory $ \widetilde{c}_{i}(t) \equiv c_{i} (t)/c_{i}^{\textrm{max}}$ of papers in Physics and Biology. 
Here, $c_{i}^{\textrm{max}}$ is the maximum number of citations received by paper $i$ in any given year after 
its publication. Fig. \ref{fig:individual} shows a summary of the
renormalization process and different measures used for analysis. For both disciplines, the citation trajectories of papers published over different years show 
systematic changes with time. New papers have higher citation rates for the first few years, whereas over longer 
periods of time old papers have higher citation rates. Some irregularity in the tail of the citation trajectories might be due to the heterogeneity in the time to reach the peak number of citations $\Delta t_{\mathrm{peak}}$. The 
change in the citation rate over time is more evident when we group the papers based on their \textit{peak year}, 
i.e., year in which they receive the maximum number of citations. Thus, the peak year represents the year in which a 
paper is at the peak of its attention. Fig.~\ref{fig:citation_trajectory} (bottom panels) show that the decay pattern is more robust 
when the papers were aggregated according to their peak year as compared to their publication year. 
This is true for other groups of papers as well: Appendix Fig.~B1 shows the same pattern for the papers in the [11-30] percentile.
\subsection{Evolution of the time to peak}
\begin{figure}[tb!]
\centering 
\includegraphics[width=0.49\columnwidth]{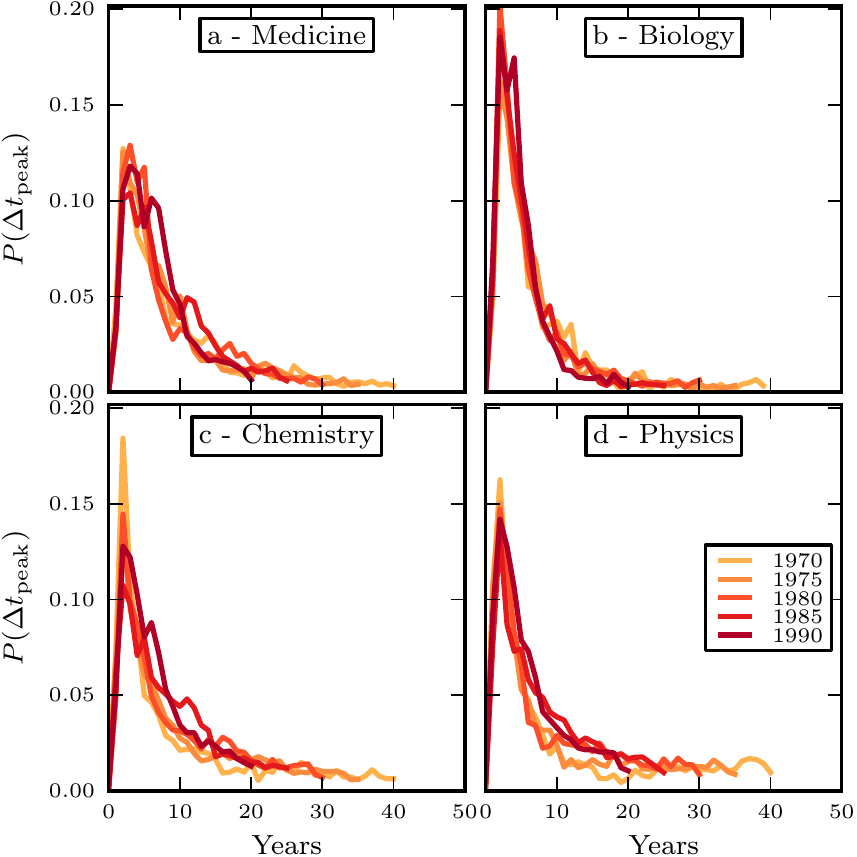}
\includegraphics[width=0.49\columnwidth]{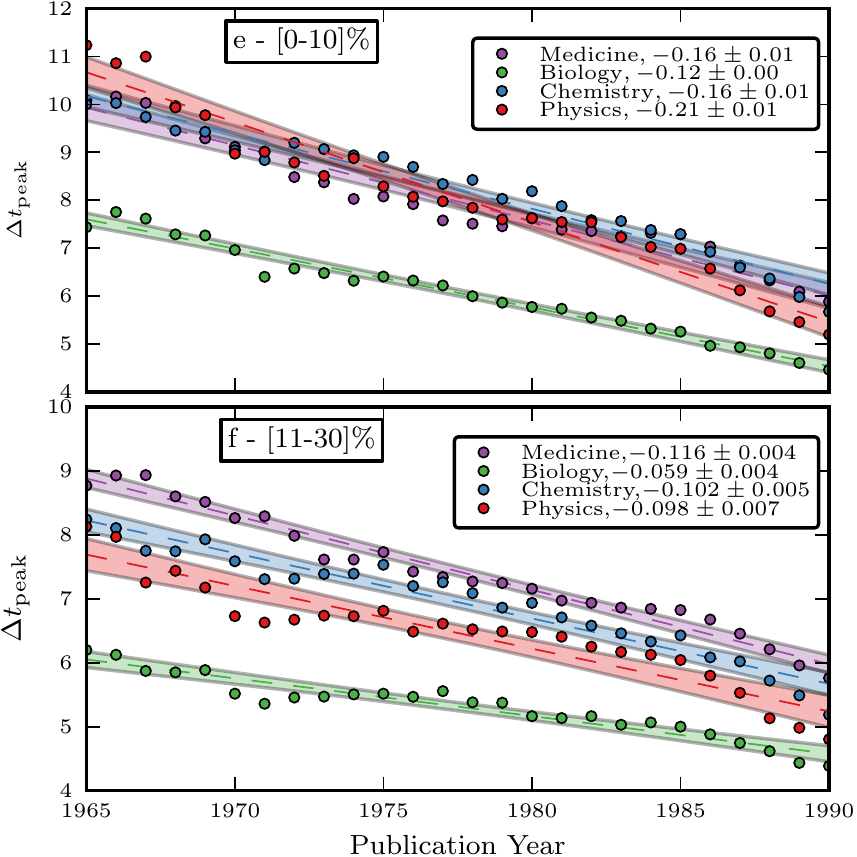}
\caption{Time to reach the peak attention $\Delta t_{\mathrm{peak}}$ is both field and time dependent. (a-d) Distribution of $\Delta t_{\mathrm{peak}}$ for papers in the top 10\% published in the same year, for different fields and publication years. (e,f) Time evolution of the mean values of $\Delta t_{\mathrm{peak}}$ for top 10$\%$ and [11-30]$\%$ percentiles. The mean value $\langle \Delta t_{\mathrm{peak}} \rangle$ decreases linearly in time. The linear fit, $95\%$ confidence interval and the slopes of the linear fits are also shown. Papers peaking after 2005 are not considered as their peak years might still be subject to change.}
\label{fig:time_to_peak}
\end{figure}

Next we investigate whether the time to reach the peak in the number of citations $\Delta t_{\mathrm{peak}}$ changes with time. In Fig.~\ref{fig:time_to_peak}~(a-d) we plot the distribution of $\Delta t_{\mathrm{peak}}$ for papers published in the same year, for all four disciplines and for several years. 
The majority of the papers peak within a few years since publication. Papers in Biology are characterized by small $\Delta t_{\mathrm{peak}}$ as compared to papers in Medicine, Physics and Chemistry. For all fields the distribution of $\Delta t_{\mathrm{peak}}$ is time dependent, with its value decreasing 
steadily in time. Fig.~\ref{fig:time_to_peak}~(e,f) shows the time evolution of the mean of $\Delta t_{\mathrm{peak}}$ for different fields and two groups of papers: the most cited 10$\%$ and the [11-30] percentile. The decreasing mean of the time to peak indicates that in recent times papers are taking less 
time to reach the peak of their attention. 
This result seems to be consistent with previous findings \citep{Egghe2010,Lariviere2008} showing, both theoretically and empirically, that the average reference age is an increasing function of time. This would suggest that more recent papers are able to dig deeper in scientific literature, reducing the amount of attention available
for papers published in recent years and therefore causing a shortening of the time needed to peak.   
Also, this behavior is shown to be independent of the citation volume of the papers, although papers with fewer citations take less time to reach the peak.  Biology shows again a unique behavior, with its values being constantly below the ones of the other fields, indicating an intrinsic faster peak time.

\subsection{Functional form of citation decay}
\begin{figure}[tb!]
\centering 
\includegraphics[width=0.49\linewidth]{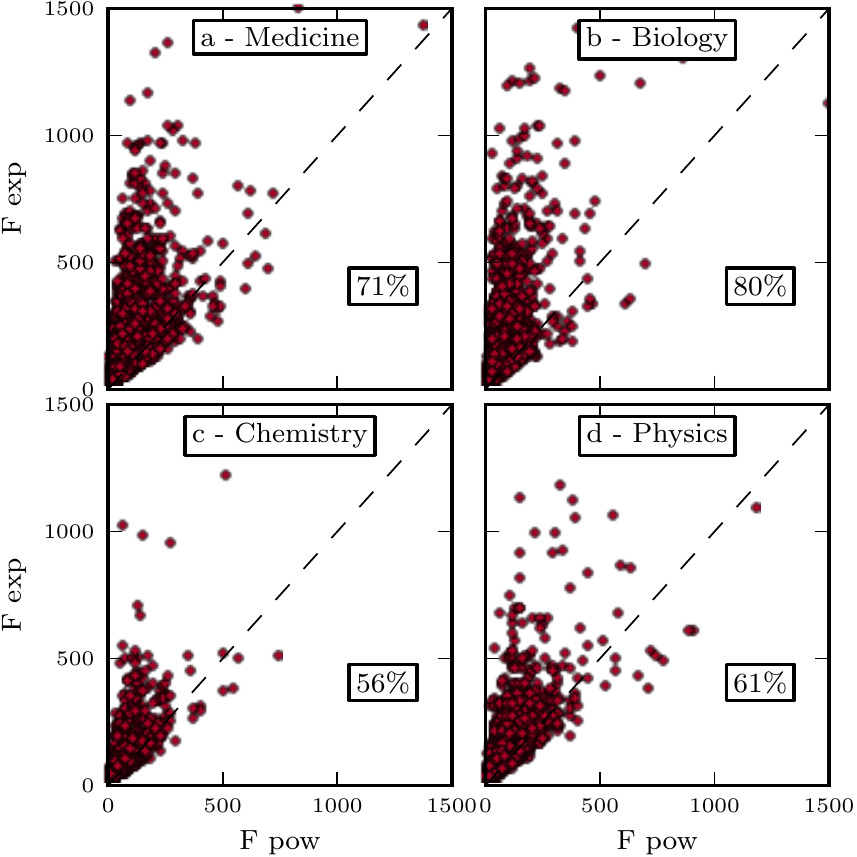}
\includegraphics[width=0.49\linewidth]{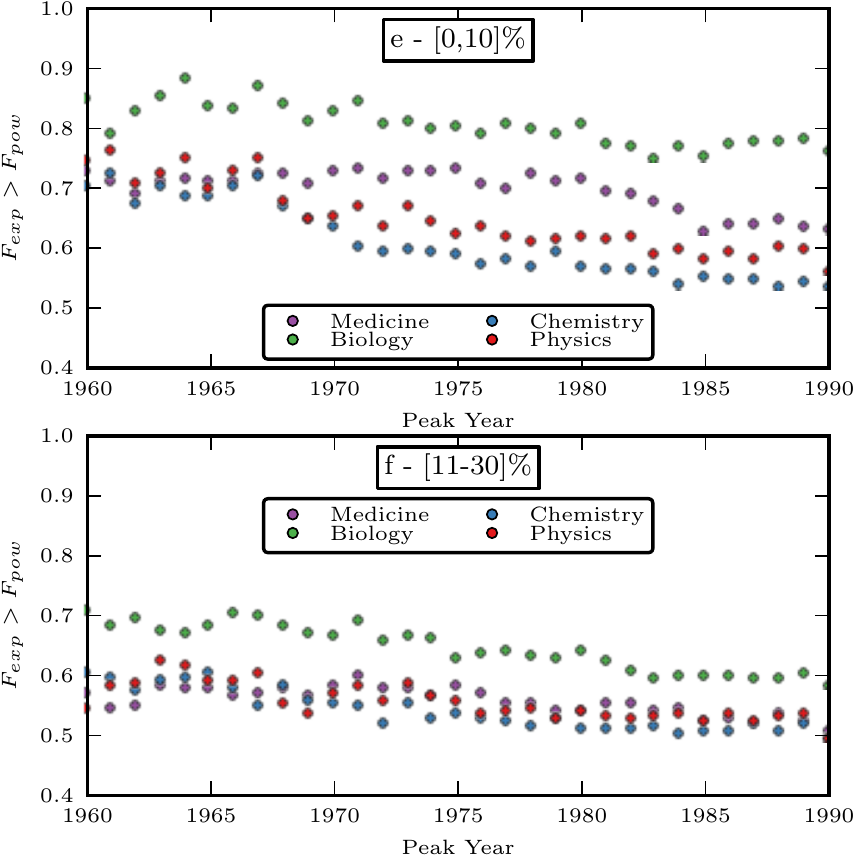}
\caption{Comparison of exponential fits with power law fits as described by the $F$-statistics. (a-d) Papers peaking in 1980, with the number in the box indicating the percentage of papers better fitted by exponentials than by power laws. In particular, it is worth noticing that there is a significant density of points in the high $F_{exp}$-low $F_{pow}$ area, showing a series of papers for which the power law fit was clearly outperformed by the exponential fit. There is no trace of the opposite scenario, with papers better fitted by power-law lying close to the diagonal line. (e, f) The time evolution of the fraction of papers for which exponentials are better descriptors than power laws, according to the F-score, for the top 10\% and [11-30]$\%$ percentiles papers over different years.}
\label{fig:f_stat}
\end{figure}

To investigate the time evolution of the change in \emph{attention} we first determine the functional form of the citation decay of each paper. We fit the normalized citation trajectories $ \widetilde{c}_{i}(t) \equiv c_{i} (t)/c_{i}^{\textrm{max}}$ using both the exponential and power law curves. We used an additional parameter in both fitting functions because the normalized citation curves after the initial decay eventually converge to a nonzero plateau. The  exponential fitting function is given by $ \widetilde{c}_{i}(t) = \beta_e \exp(- \alpha_e t) + \gamma_e $ whereas the power law fitted function is given by  $ \widetilde{c}_{i}(t) = \beta_p t^{-\alpha_p}  + \gamma_p$. We fit the normalized citation trajectories of each paper and determine the best fit parameters using the least square method. First, we found that for the majority of the papers both the exponential and power law decrease could fit the decaying behavior, since the $p$-value of the fit is less than $10^{-3}$. However, comparing the two fits for each paper using F-statistics, we found that the exponential fits better the decaying behavior. Fig.~\ref{fig:f_stat} shows that for most paper $F$-statistics is much larger for the exponential fit as compared to the power law fit. Interestingly, in recent years the fraction of papers that fits a power-law curve has been increasing systematically.  
Fig.~\ref{fig:f_stat}~(e) shows the time evolution of the fraction of papers whose F-score in the exponential fitting exceeds the F-score for the power law case for the top 10\% decile. All the four fields show a trend where the power law fit gradually improves in time. This phenomenon may be linked to the smaller impact of the convergence to the final plateau, on the fit.  On average the convergence to the plateau takes more than 20 years, and papers in recent years might not have reached this plateau in their decay.

\subsection{Ultradiffusion and decay in attention}
A trademark of the evolution of the number of citations of a paper is their decline after reaching a peak. Here, we provide an explanation of this decay. Each citation is considered an \emph{event} and the temporal evolution of the number of citations (after the peak) is taken as a \emph{counting process}. The observed counting process could be rationalized as ultradiffusive if it has signatures associated with an ultradiffusive process. Ultradiffusion is a stochastic process where every timestamp of a  timeseries  $\{t_i\}$ ($t_i<t_j$ if $i<j$) $\forall i \in 0\cdots n$ is associated with an event $\{X_{t_n-t_i}\}$. State $X_{t_n-t_0}$ is analogous to the event of citing the paper. All the other states are associated with not citing the paper.  Unlike the Poisson process, which assumes that events occur independently of each other, ultradiffusion elicits that a later event might be caused by or correlated to an earlier event or a combination of earlier events. The earlier event in turn might be independent or might be correlated to a combination of even earlier events. This leads to a hierarchical causal/correlational model of prior event occurrences which can be used to predict the occurrence of a new event. Thus, ultradiffusion proposes that the observed pattern of events is a consequence of an  underlying hierarchy of states. In this hierarchical model, an event temporally nearer to the occurring event has a greater probability of affecting it. In other words, the correlation between two events is determined by a notion of ``closeness'' or distance between them.

For any ultradiffusive process there must be an ultrametric space on which distances between occurrences are defined. In this case the distance between two events $X_{t_i}$ and $X_{t_j}$ can be defined as 
\begin{equation}
d(X_{t_i},X_{t_j}) = \begin{cases}
  |max(t_n-t_i, t_n-t_j)| , & \text{if } i \ne j, \\
  0, & \text{otherwise}.
\end{cases}
\label{eq:ultrametric}
\end{equation}
The above definition of distance satisfies the \emph{ultrametric distance metric properties} because:
\begin{enumerate}
\item $d(X_{t_i},X_{t_j}) \ge 0$ (non-negative)
\item  $d(X_{t_i},X_{t_j})=0$ if $i=j$ (identity of indiscernibles)
\item $d(X_{t_i},X_{t_j})=d(X_{t_j},X_{t_i})$ (symmetry)
\item $d(X_{t_i},X_{t_j})\le max(d(X_{t_i},X_{t_k}),d(X_{t_k},X_{t_j}))$ (ultrametric property).
\end{enumerate}
Therefore the associated space is ultrametric~\citep{ghosh14}.
For an unltradiffusive process, the autocorrelation $P_{X_{t_i}}(t)$ , i.e., the probability of finding the system at the initial state $X_{t_i}$ after time $t$ can be calculated analytically. The autocorrelation function has an exact solution for an ultrametric space defined by a hierarchical tree. 
Assuming that the rate of transition between states is $X_{t_i}$ and $X_{t_j}$ is $e^{-\mu d(X_{t_i},X_{t_j})}$ and the probability of citing the paper is 1 when  the peak in the number of citations is reached,  the probability of citing the paper at time $t$ is given by $P_{X_{t_n-t_0}}(t)$. 
When the number of states is finite, such an autocorrelation function is exponential in nature, otherwise it follows a power law behavior ~\citep{BachasHuberman}.

\subsection{Evolution of the decay exponent}
\begin{figure}[tb!]
\centering 
\includegraphics[width=0.49\columnwidth]{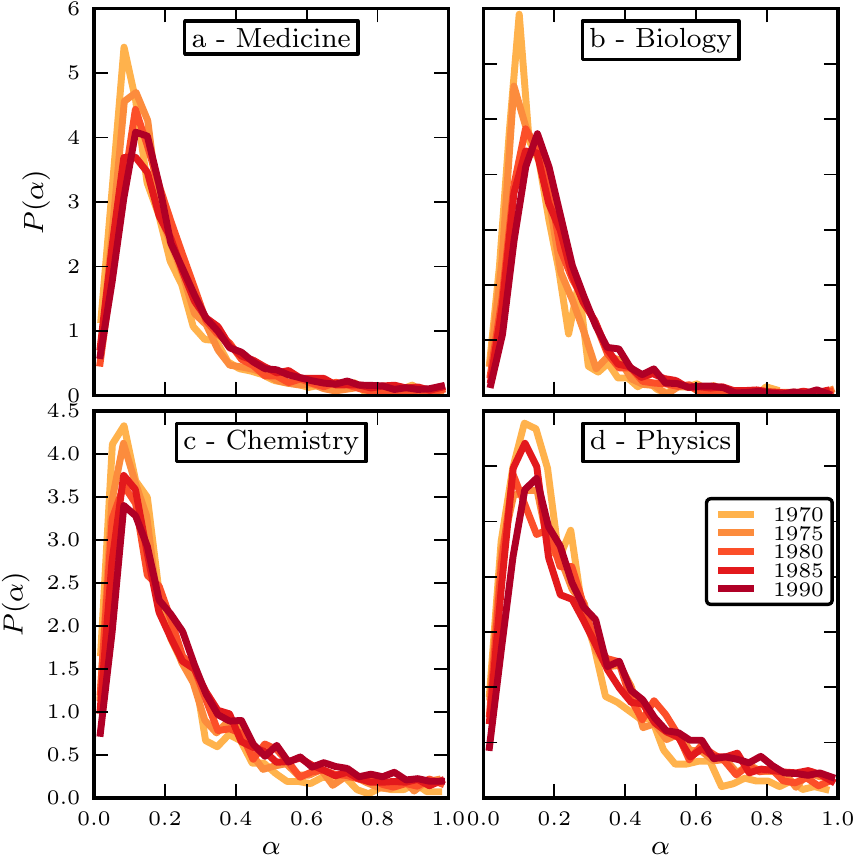}
\includegraphics[width=0.49\columnwidth]{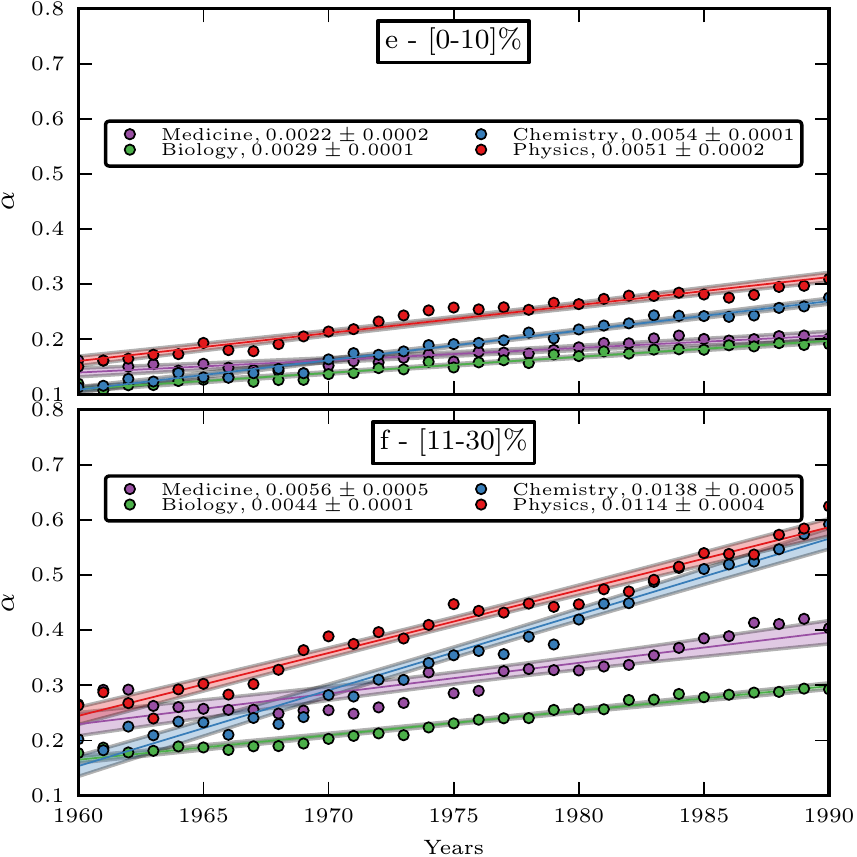}
\caption{Attention to publication is decaying faster in time. (a-d) Distribution of parameter $\alpha$ for exponential fits in different years for the four disciplines. For recent years the tail of the distribution becomes progressively fatter. (e-f) Time evolution of the median of the distributions of the decay rates $\alpha$, along with linear fit, $95\%$ confidence interval and slopes.  The top panel refers to the top $10\%$ most cited papers, the bottom panel to the [11-30] percentile. The data suggests a ``grouping'' of Medicine and Biology vs Physics and Chemistry, with the two groups having nearly identical numbers for the fit. Moreover, for the [11-30] range the coefficients are nearly doubled compared to [0-10]. This means that the speed of the decay depends on the citation volume of each paper.}
\label{fig:exp_fit}
\end{figure}

Fig.~\ref{fig:exp_fit} shows the distributions of the exponential decay rates $\alpha_e$ for papers grouped by their peak years. The distributions for different disciplines show that majority of papers have a characteristic rate. Moreover, for all the disciplines the shape of the distribution is broader for papers peaking in recent years. The median of the distributions shows a systematic increase in time (Fig.~\ref{fig:exp_fit}~e,f). Such a faster decay behavior is independent of the fitting ansatz. Furthermore, this pattern is independent of the group of papers chosen for the analysis (top 10$\%$ for top panel, [11-30] percentile for bottom panel). This suggests that the later a paper peaks, the shorter is its life cycle, implying a faster decay of scientific attention in terms of absolute time.
The decay rates and their relative increase with time appears to be field dependent. For example, for Physics and Chemistry the decay is faster compared with Biology and Medicine. 

\subsection{Exponential increase in number of publications}
\begin{figure}[tb!]
\centering 
\includegraphics[width=0.5\linewidth]{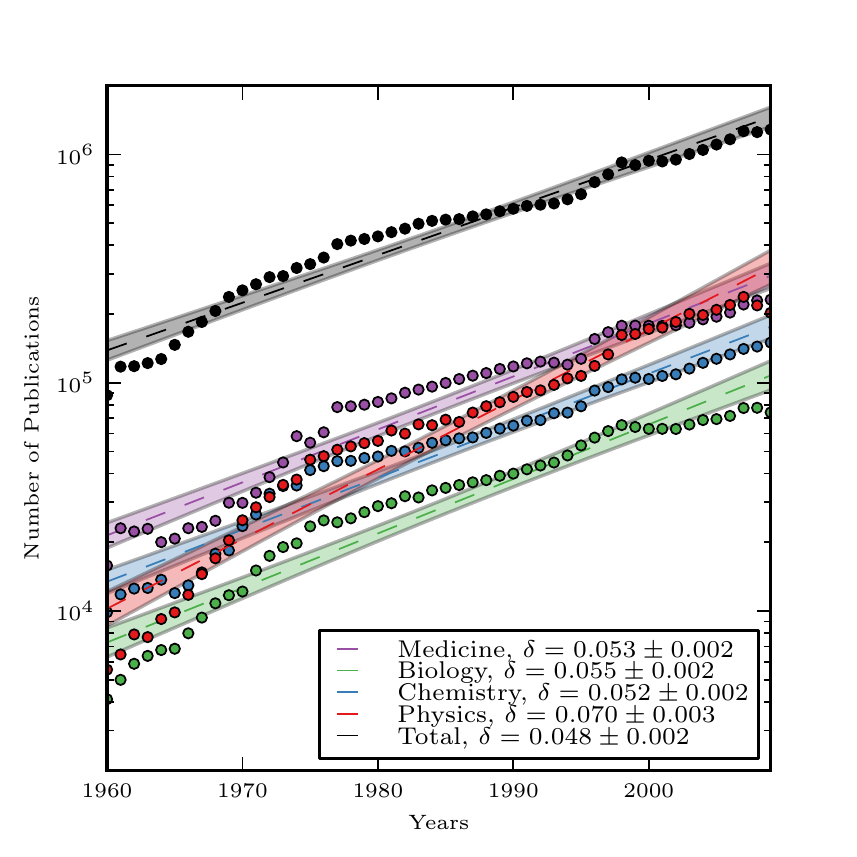}
\caption{Increase in the number of publications with time since 1960 along with exponential fits, $95\%$ confidence intervals and rates.}
\label{fig:nPublications}
\end{figure}

The progressively faster decay in attention we observe is compatible with the intuitive picture of scientific theories and papers constantly replaced by other competing results. As the number of publications is also growing with time, it takes less time to replace or update older scientific results. Thus, the rapid increase in the number of papers could provide an explanation. In Fig.~\ref{fig:nPublications} we report the growth of the number of publications in different fields with time, fitted by the function $N_{p} = N_{0} \exp^{\delta t}$. All the fields show an exponential increase, as observed for the total number of publications.

Hence, the process of attention gathering needs to take into account the increasing competition between scientific products. With the increase of the number of journals and increasing number of publications in each journal (not to mention the growth of online journals, which do not have physical constraints in their publication volume), a scientist inevitably needs to filter where to allocate its attention, i.e. which papers to cite, among an extremely broad selection. This may also question whether a scientist is actually fully aware of all the relevant results available in scientific archives. Even though this effect is partially compensated by the increase of the average number of references, one needs to consider the impact of increasing publication volume on the attention decay.

\begin{figure}[htb!]
\centering 
\includegraphics[width=0.49\linewidth]{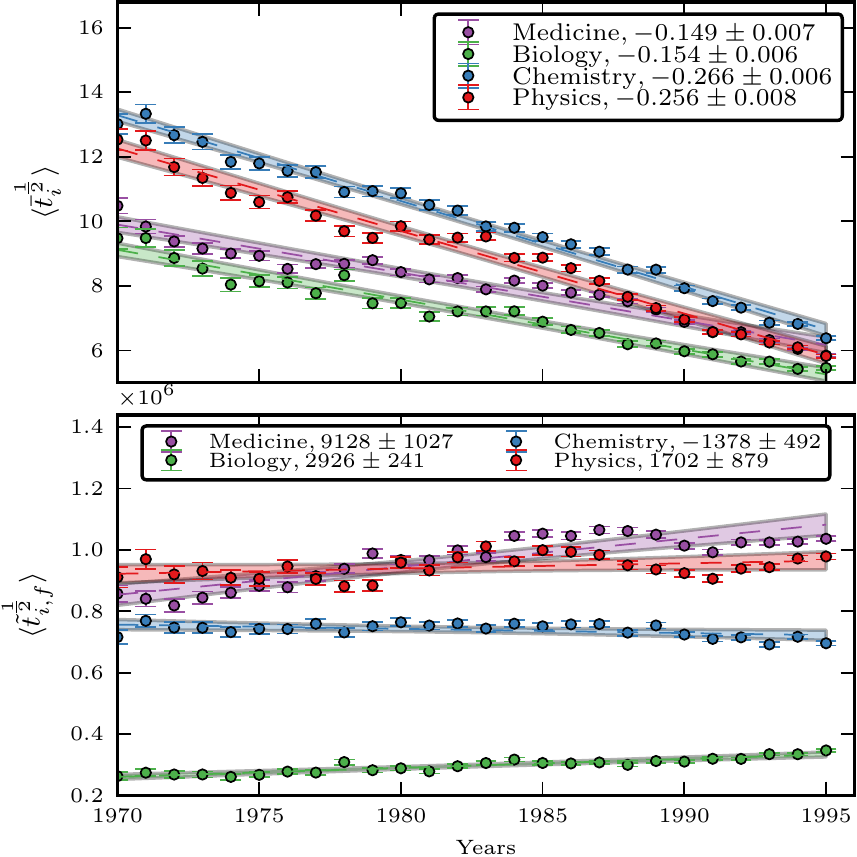}
\includegraphics[width=0.49\linewidth]{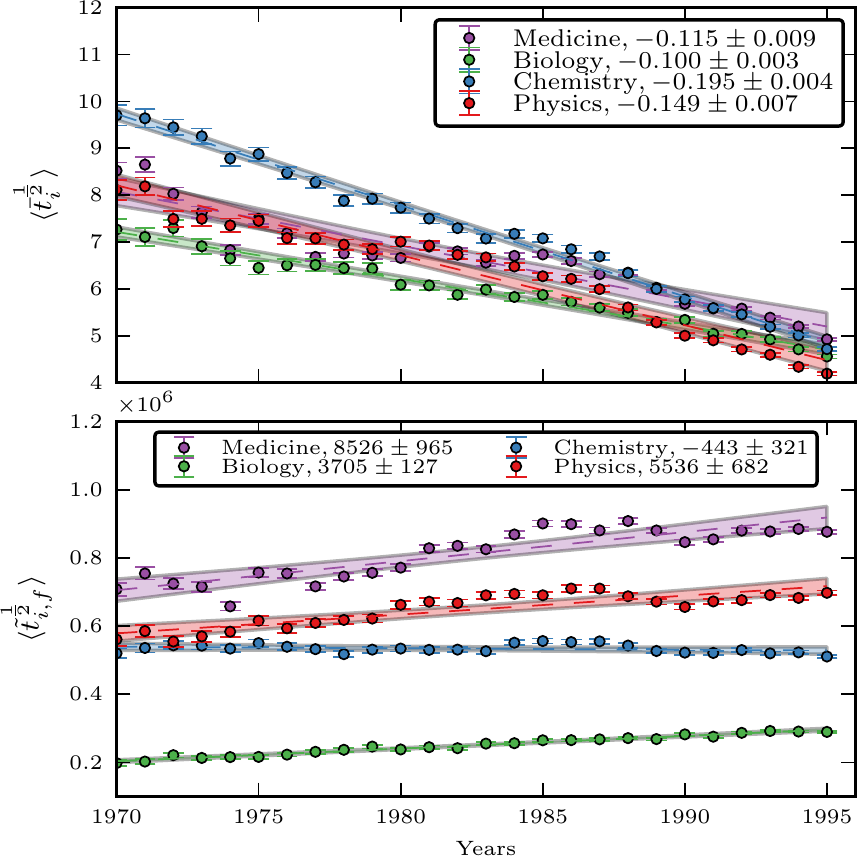}
\caption{The half-life of papers $t_{i}^{\frac{1}{2}}$ in terms of absolute time decreases linearly, whereas the rescaled half-life of papers $\widetilde{t}^{\frac{1}{2}}_{i}$ in terms of the number of publications is relatively constant. The panels show the evolution of $\langle t_{i}^{\frac{1}{2}} \rangle$ (top) and $\langle \widetilde{t}_{i,f}^{\frac{1}{2}} \rangle$ (bottom) for the four different fields and for the top 10$\%$ (left) and for the [11-30]$\%$ percentile (right). The latter values are divided by a large constant to get small values on the y-axis, which are easier to display. The error bars indicate the standard errors. Linear fits along with their $95\%$ confidence level intervals are also shown. In the legend the values of the linear coefficients are shown for both absolute ($q_{y}$) and renormalized ($q_{r}$) time. The dashed line represents the linear fit. Despite its noisy behavior, the renormalized half-life shows a relatively stable trend throughout the years, possibly with the only exception of Medicine and Biology, which show a slightly rising pattern for recent times.}
\label{fig:HalfLife}
\end{figure}
\subsection{Half-life}
To check the robustness of our result that the citation decay rate is becoming faster for recent papers, we measure the \textit{half-life} of each publication. The \textit{half-life} of a paper is a metric regularly adopted to evaluate the typical life-cycle of a paper. The \textit{half-life} of a paper is the time after which the normalized citation rate $\widetilde{c}_{i}(t)$ is never above $\frac{1}{2}$. Similarly, instead of $1/2$, other thresholds $\sigma$ of the citation rate can also be considered. In mathematical terms:
\begin{equation}
  t_{i}^{\frac{1}{2}} =  \max \{ t \hspace{2mm} \mathrm{s.t.} \hspace{2mm}  \widetilde{c}_{i}(t) \geq \frac{1}{2} \}.
\end{equation}
The value $t^{\frac{1}{2}}_{i}$ is the year of the last ``sub-peak'' of attention for paper $i$ as it quantifies the last moment in the history of the paper at which it has been able to gather sufficient attention.
Fig.~\ref{fig:HalfLife}(top panels) shows the time evolution of the half-life measure. The mean of the absolute measure $\langle t^{\frac{1}{2}}_{i} \rangle$ decreases linearly with time for all the four fields. This decrease is consistent with the linear increase in the decay rate of the citation trajectory. Also, there is an interesting grouping between Medicine/Biology and Chemistry/Physics: they start off widely separated but they converge pairwise to similar values in recent years. 

\subsection{Rescaling time}
The \textit{half-life} of a paper can also be used to analyze the impact of the growth of system size. Using the data shown in Fig.~\ref{fig:nPublications}, we are able to convert its value from a measure of time into a measure of number of publications in the paper's discipline that have been published between the peak of the paper and $t^{\frac{1}{2}}_{i}$. Therefore we are able to define a renormalized version of $t^{\frac{1}{2}}_{i}$ as:
\begin{equation} 
  \widetilde{t}^{\frac{1}{2}}_{i,f} = \sum_{t = t^{\textrm{peak}} + 1}^{t^{\frac{1}{2}}_{i}} N_{p}^{f}(t)
  \label{eq:renormalization}
\end{equation} 
where $t^{\textrm{peak}}$ stands for the peak year and $N_{p}^{f}(t)$ indicates the number of publications in field $f$ of paper $i$ for year $t$.

Fig.~\ref{fig:HalfLife}(bottom panels) shows the time evolution of the renormalized half-life measure. Contrary to the previous measure, the evolution of the renormalized half-life $\langle \widetilde{t}^{\frac{1}{2}}_{i,f} \rangle$ shows a relatively stable behavior. Note that, this observation is highly non-trivial as the stable renormalized half-life is only expected in the case when the exponential increase in the number of publications exactly compensate for the decay in citation rate.    
A similar behavior is also observed when lower thresholds $\sigma$ are used, i.e., by forcing the drop to be more significant (see Appendix Fig.~C2~a). The renormalized half life defined in equation \ref{eq:renormalization} provides a measure of the time required for a paper to fall below a certain arbitrarily defined threshold of attention in terms of number of publications, which can be seen to represent the amount of "competition" a paper is about to withstand before dropping to significantly lower values of attention. 

Interestingly, the picture changes if we consider the half-life to be the first time when the normalized citation rate $\widetilde{c}_{i}(t)$ decreases below $\frac{1}{2}$. In this case, the renormalized half-life shows an increasing pattern with time (Appendix Fig.~C2~b). Such alternative measure quantifies the time taken to have the first lowest drop of attention. However data suggests that such value seems to be stable across years for each field as an initial drop in attention appears to be structurally inevitable. This inevitably leads, after renormalization, to a significantly increasing behaviour. 

Fig.  ~\ref{fig:HalfLife} suggests that, even though papers are now taking on average less time to drop below a certain threshold of attention, the number of published papers after which a work becomes obsolete does not show the same behavior. On the contrary, our data indicates an approximately constant value throughout the time period of the study. 
So, the growing number of publications proportionally increases the likelihood of a paper to become obsolete, but the contribution of each paper to this process is about the same, regardless of the age of the paper.

\section{Conclusions}

We have studied how attention towards scientific publications diminishes over time, due to the obsolescence of knowledge. For millions of papers in four different disciplines we find that
after reaching a peak, typically a few years since publication, the number of citations goes down relatively fast. We find that exponential decays 
are to be generally preferred over power law decays, though the latter are providing better and better descriptions of the data for recent times. 
 The existence of many time-scales in citation decay and our ability to construct an ultrametric space to represent this decay, leads us to speculate that citation decay is an ultradiffusive process, like the decay of popularity of online content.
Interestingly, the decay is getting faster and faster, indicating that scholars ``forget" more easily papers now than in the past. We found that this has to do with the 
exponential growth in the number of publications, which inevitably accelerates the turnover of papers, due to the finite capacity of scholars to keep track of the scientific literature. Although search engines and digitalization have made it easier for scientists to discover relevant information, the amount of information that can be successfully processed is still limited.   
In fact, by measuring time in terms of the number of published works, the decay appears approximately stable over time, across disciplines, although there are slight monotonic trends for Medicine and Biology. However, we must emphasise that we normalized time by using the number of published papers 
in the discipline at study. This is the simplest choice to make, but it is not necessarily the most sensible one. The fields we considered are rather broad, and subdivided in many 
different topics. Scholars working on any of such topics will be affected mostly by the literature of the topic, and hardly by anything else. It is very difficult to isolate the relevant literature case by case. Still, considering the whole bulk of publications in each single discipline is a way to discount the exponential growth of scientific output and we have found that this suffices to 
counterbalance (at least to a large extent) the apparent faster decay of attention observed in recent years.

\section*{Acknowledgments}
We used data from the Science Citation Index Expanded, Social Science Citation Index and Arts \& Humanities Citation Index, prepared by Thomson Reuters, Philadelphia, Pennsylvania, USA, Copyright Thomson Reuters, 20
. We gratefully acknowledge KNOWeSCAPE, COST Action TD1210 of the European Commission, for fostering interactions with leading experts in science of science who gave feedback on the paper. We also thank HP Labs for supporting the visit of SF, during which the project 
was started.

\section*{Author Contributions}
All authors designed the research and participated in the writing of the manuscript. PDBP and RKP collected and analysed the data. PDBP performed the research.

\section*{References}

\appendix
\setcounter{figure}{0}    
\setcounter{table}{0}    

\section{Description of the categories}

To categorize each paper according to its field of publication we use the Thomson Reuters (TR) subject categories. We then aggregated these subject categories into broader scientific fields. A detailed description is provided in Table~\ref{tab:cat}

\begin{center}
\begin{table}[t!]
  \begin{tabular}{| p{3.8cm} | >{\footnotesize}p{12cm} |}
\hline
Fields & TR subject categories \\ 
\hline
Physics & IMAGING SCIENCE \& PHOTOGRAPHIC TECHNOLOGY;
 PHYSICS, APPLIED;
 OPTICS;
 INSTRUMENTS \& INSTRUMENTATION;
 PHYSICS, CONDENSED MATTER;
 PHYSICS, FLUIDS \& PLASMAS;
 PHOTOGRAPHIC TECHNOLOGY;
 PHYSICS, ATOMIC, MOLECULAR \& CHEMICAL;
 ACOUSTICS;
 PHYSICS;
 PHYSICS, MATHEMATICAL;
 MECHANICS;
 PHYSICS, NUCLEAR;
 SPECTROSCOPY;
 THERMODYNAMICS;
 PHYSICS, PARTICLES \& FIELDS;
 NUCLEAR SCIENCE \& TECHNOLOGY;
 PHYSICS, MULTIDISCIPLINARY;
 ASTRONOMY \& ASTROPHYSICS;
\\ \hline
Chemistry & CHEMISTRY, INORGANIC \& NUCLEAR;
 ELECTROCHEMISTRY;
 CHEMISTRY, PHYSICAL;
 CHEMISTRY, ANALYTICAL;
 POLYMER SCIENCE;
 CHEMISTRY, MULTIDISCIPLINARY;
 CRYSTALLOGRAPHY;
 CHEMISTRY, APPLIED;
 CHEMISTRY;
 CHEMISTRY, ORGANIC;
\\ \hline
Molecular Biology & BIOCHEMICAL RESEARCH METHODS;
  BIOCHEMISTRY \& MOLECULAR BIOLOGY;
  BIOMETHODS;
  BIOPHYSICS;
  CELL \& TISSUE ENGINEERING;
  CELL BIOLOGY;
  CYTOLOGY \& HISTOLOGY;
  MATHEMATICAL \& COMPUTATIONAL BIOLOGY;
  MICROSCOPY;
\\ \hline
Physiology or Medicine & CYTOLOGY \& HISTOLOGY;
  BIOCHEMISTRY \& MOLECULAR BIOLOGY;
  CELL BIOLOGY;
  BIOCHEMICAL RESEARCH METHODS;
  CELL \& TISSUE ENGINEERING;
  MATHEMATICAL \& COMPUTATIONAL BIOLOGY;
  BIOPHYSICS;
  BIOMETHODS;
  MICROSCOPY;
  ENGINEERING, BIOMEDICAL;
  IMMUNOLOGY;
  MEDICAL LABORATORY TECHNOLOGY;
  MEDICINE, RESEARCH \& EXPERIMENTAL;
  PARASITOLOGY;
  PHYSIOLOGY;
  ANATOMY \& MORPHOLOGY;
  PATHOLOGY;
  ONCOLOGY;
  RHEUMATOLOGY;
  VASCULAR DISEASES;
  PSYCHIATRY;
  GERIATRICS \& GERONTOLOGY;
  DENTISTRY, ORAL SURGERY \& MEDICINE;
  OPHTHALMOLOGY;
  DENTISTRY ORAL SURGERY \& MEDICINE;
  MEDICINE, LEGAL;
  EMERGENCY MEDICINE \& CRITICAL CARE;
  CLINICAL NEUROLOGY;
  TRANSPLANTATION;
  HEMATOLOGY;
  INFECTIOUS DISEASES;
  RESPIRATORY SYSTEM;
  PERIPHERAL VASCULAR DISEASE;
  MEDICINE, GENERAL \& INTERNAL;
  PEDIATRICS;
  EMERGENCY MEDICINE;
  INTEGRATIVE \& COMPLEMENTARY MEDICINE;
  GASTROENTEROLOGY \& HEPATOLOGY;
  DERMATOLOGY;
  REHABILITATION;
  ANESTHESIOLOGY;
  TROPICAL MEDICINE;
  MEDICINE, MISCELLANEOUS;
  ENDOCRINOLOGY \& METABOLISM;
  NEUROIMAGING;
  ANDROLOGY;
  ORTHOPEDICS;
  OBSTETRICS \& GYNECOLOGY;
  ALLERGY;
  CRITICAL CARE MEDICINE;
  OTORHINOLARYNGOLOGY;
  RADIOLOGY, NUCLEAR MEDICINE \& MEDICAL IMAGING;
  SURGERY;
  CARDIAC \& CARDIOVASCULAR SYSTEMS;
  DERMATOLOGY \& VENEREAL DISEASES;
  AUDIOLOGY \& SPEECH-LANGUAGE PATHOLOGY;
  RADIOLOGY \& NUCLEAR MEDICINE;
  UROLOGY \& NEPHROLOGY;
  CRITICAL CARE;
  CARDIOVASCULAR SYSTEM;
\\ \hline
\end{tabular}
\caption{Aggregation of TR subject categories in broader fields.}
\label{tab:cat}
\end{table}
\end{center}

\section{Evolution of the number of citations for other decile}
\begin{figure}[htb!]
\centering 
\includegraphics[width = 0.5\linewidth]{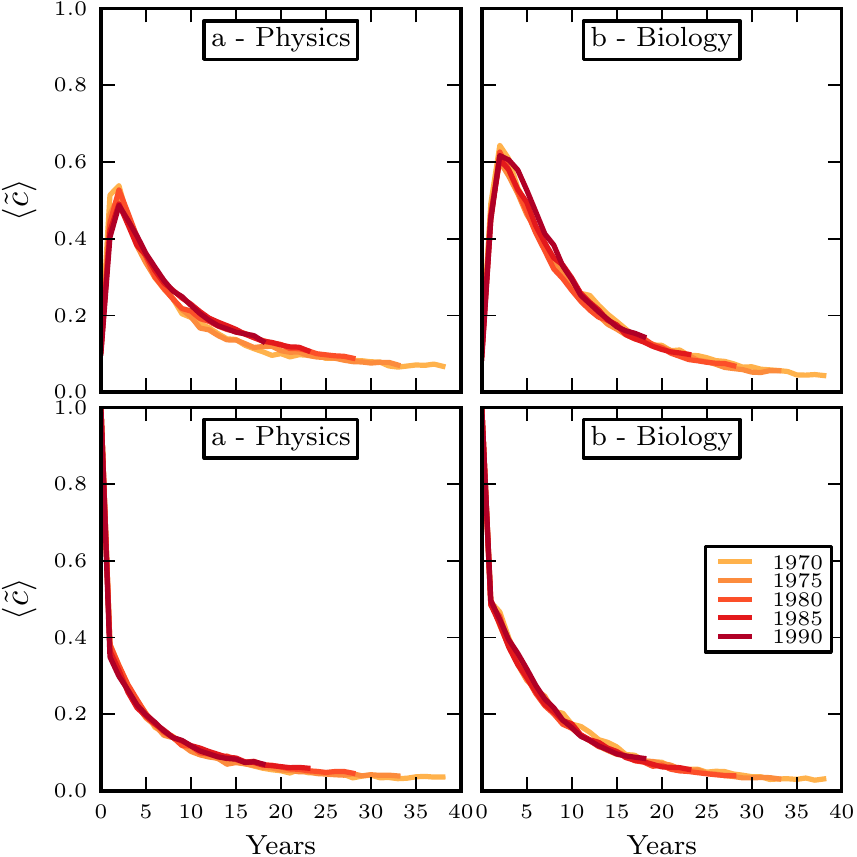}
\caption{Averaged citation trajectories are calculated for papers in the  [11-30]\% window based on their total number of citations.}
\label{fig:citation_trajectory_11_30}
\end{figure}

Fig.~\ref{fig:citation_trajectory_11_30} is the analog of figure Fig.~\ref{fig:citation_trajectory} of the main text, but is focused on the top [11-30]\% papers (based on their total number of citations). Compared to the original figure the values of $\langle  \widetilde{c} (t) \rangle$ is lower, linked to the fact that these papers have accumulated fewer citations. The top panels (A,B), where the papers are grouped by their publication year, show that the average peak is more concentrated in the initial years and is followed by a more rapid decay. Finally, the citation trajectories reach a plateau that is significantly lower than the respective one for the top decile. Similarly, the papers grouped by their peak year (bottom panels, C,D), also show a larger drop in $\langle  \widetilde{c} (t) \rangle$ in the first few years followed by a lower value of the final plateau.

\section{Evolution of half-life for different values of $\sigma$ and alternative definition of half-life}

\begin{figure}[htb!]
\centering 
\includegraphics[width = 0.49\linewidth]{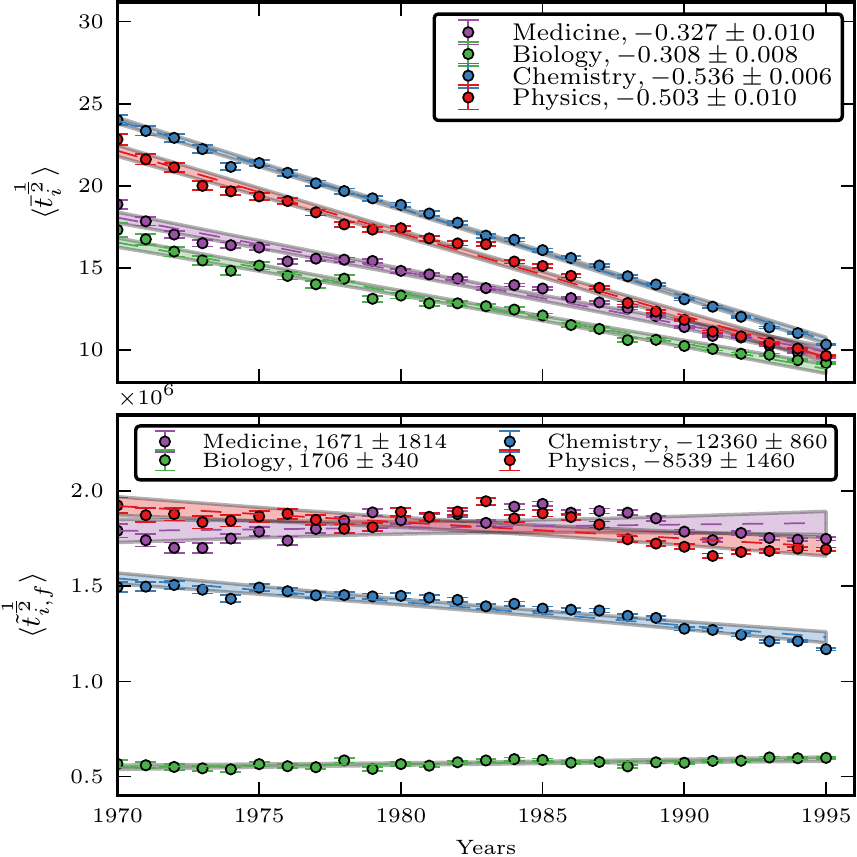}
\includegraphics[width = 0.49\linewidth]{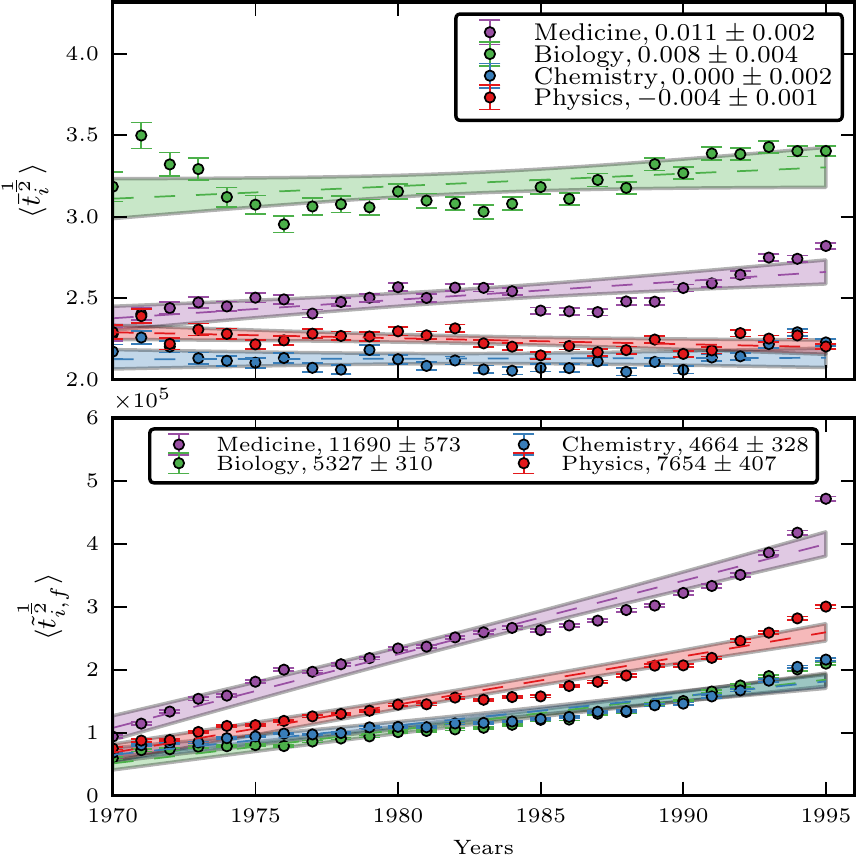}
\caption{(Left) The half-life of papers $t_{i}^{\frac{1}{2}}$ with $\sigma = 0.3$. (Right) The alternative half-life of papers $\bar{t_{i}}$ with $\sigma = 0.5$. }
\label{fig:HalfLife_0.3}
\end{figure}

Fig.~\ref{fig:HalfLife_0.3}~(a,b) is the analogous of Fig.~\ref{fig:HalfLife} with $\sigma = 0.3$. This implies choosing a lower threshold for the definition of the point below which a paper is considered to have completed its life cycle. Data suggests that the pattern shown in the paper is retained for other choices of parameters. However, at $\sigma=0.3$, Physics also shows a slight decreasing patter, whereas Medicine and Biology retain their increasing trends. 

Fig.\ref{fig:HalfLife_0.3}~(c,d) is the analog of the previous figure, with the alternative half-life defined as 
\begin{equation}
 \bar{t_{i}}^{\frac{1}{2}} =  \min \{ t \hspace{2mm} \mathrm{s.t.} \hspace{2mm}  \widetilde{c}_{i}(t) \leq \frac{1}{2} \}.
\end{equation}
whereas $\bar{t}$ is defined still in the same way as in Eq.~\ref{eq:renormalization} but using the previously defined value for $t$.  In th framework the half life of the paper is considered as the first year in its life cycle where its citations have dropped below a certain threshold. The figure shows that with this definition the values of $\bar{t}$ lose their decreasing pattern in favour of a field specific value, which is retained in the years. Similarly, the behavior for $\bar{t}$ shows a deviation from the previously constant pattern in favor of a significant increase in its values.

\end{document}